\documentclass[a4paper]{llncs}
\usepackage{graphicx, wrapfig, float, hyperref, xcolor}
\usepackage[numbers,sort&compress]{natbib}
\graphicspath{{images/}}

\makeatletter 
\renewcommand\@biblabel[1]{#1} 
\makeatother
\begin{document}
\title{Artificial Neural Network Surrogate Modeling of Oil Reservoir: a Case Study}
%
%
\author{Oleg Sudakov\inst{1} \and
Dmitri Koroteev\inst{1} \and
Boris Belozerov\inst{2} \and
Evgeny Burnaev\inst{1}}
\authorrunning{O. Sudakov, D. Koroteev, B. Belozerov, E. Burnaev}
%
\institute{Skolkovo Institute of Science and Technology, Moscow, Russia \and
Gazprom Neft, Science and Technology Center, St. Petersburg, Russia}
\maketitle              
\begin{abstract}
We develop a data-driven model, introducing recent advances in machine learning to reservoir simulation. We use a conventional reservoir modeling tool to generate training set and a special ensemble of artificial neural networks (ANNs) to build a predictive model. The ANN-based model allows to reproduce the time dependence of fluids and pressure distribution within the computational cells of the reservoir model. We compare the performance of the ANN-based model with conventional reservoir modeling and illustrate that ANN-based model (1) is able to capture all the output parameters of the conventional model with very high accuracy and (2) demonstrate much higher computational performance. We finally elaborate on further options for research and developments within the area of reservoir modeling.
\end{abstract}

\section{Introduction and Justification}
Hydrodynamical reservoir modeling based on variations of Darcy’s law has been an acknowledged foundation for field development for several decades \citep{chavent1986mathematical,tyson2007introduction,nguyen2012capacitance}. This modeling is typically based on numerical integration of diffusion-type equations within a computational domain represented by several hundreds to several tens of billions of cells. The number of cells depends on the complexity of reservoir geology, field development system and placement of the wells. Some of the modern reservoir simulation software packages support multi-CPU and sometimes multi-GPU environments, which allow to speed up the computations. But even highly parallelized software releases can execute a single field development scenario for several days and even weeks for complex reservoir systems. 

We strive to build models for fast approximation, or the so-called surrogate models, to estimate selected properties based on the results of physical modeling (for details, see example of surrogate model application in \citep{belyaev2016gtapprox,HDA2013}). Surrogate models are a well-known way of solving various industrial engineering problems \citep{grihon2013surrogate,SMspacecraft2016, Grihon2014, GrihonFactorial2014, drilling2019,oil2019,core2019,makhotin}. 

Previous efforts on surrogate modeling for reservoir development could be found in \citep{lasdon2017implementing,mohaghegh2000design,pyrcz2014geostatistical,mohaghegh2012grid,kalantari2015coupling,mohaghegh2014converting}. These approaches were able to predict either well-by-well or a field scale production. Our approach differs significantly from the ones presented in \citep{lasdon2017implementing,mohaghegh2000design,pyrcz2014geostatistical,mohaghegh2012grid,kalantari2015coupling,mohaghegh2014converting}. We aim to predict the evolution of distribution of pressures and reservoir fluids within all the computational cells.

Therefore, in this paper we consider utilization of recent advances in deep Artificial Neural Network based models (ANNs) to speed up the simulations by substituting specific time consuming portions of physics-driven simulation by relatively fast data-driven forecast.

In Section \ref{sec1} we describe the initial physical simulator, used ANN architecture and its training. In Section \ref{sec2} we provide results of computational experiments. In Section \ref{sec3} we draw conclusions and discuss possible future research directions.

\section{Methods}
\label{sec1}
\subsection{Description of Features}

The information about the state of a hydrocarbon reservoir is typically stored for numerous time steps. It might contain either time series of actual records of fluid production from the wells or simulated data with a fixed time resolution. The latter contains distribution of pressure and saturation of reservoir fluids over 3D cells, which represent the reservoir model. The pressure and saturation values are considered to be constant within a cell for a given time step. These values are obtained using conventional mathematical modeling software, such as Eclipse, TNavigator or CMG software products \citep{eclipse,cmg,tnavigator}.

Depending on several factors, a cell might be either active or inactive. An inactive cell does not play significant role in petrophysical processes in the reservoir and is disregarded both in conventional reservoir modelling and in surrogate modelling procedure. The same cell can be active (contain a numerical value) for some of the physical characteristics, and be inactive for others. For example, a given cell might simultaneously have pressure value (be active for pressure) and lack value for another physical characteristic (be inactive for it).

In this study we use the reservoir model represented by a parallelogram computational domain containing $70 \times 145 \times 114$ cubic cells. The number of active cells for the most of physical characteristics is 715112. This amount makes straightforward application of conventional machine learning techniques, such as linear regression, or decision trees \cite{Hastie2010} relatively infeasible even for two sequential timesteps, due to the size of the sample vector and of complete dataset. 3D convolutions, which were explicitly designed to work with similar data \citep{notchenko2017large}, are impractical for the same reason when using conventional workstations. The total number of timesteps exceeds 100, which also complicates the usage of conventional algorithms.

The information about oil reservoir for a given timestep is not limited to 3D cubic structure. For each timestep, we have a set of directives for every oil well. For our purposes, we considered three kinds of directives: oil extraction, water extraction and water injection. Depending on the timestep, these can be historical values or the ones, acquired with petrophysical mathematical modelling software. Generally, extraction of the coordinates of each oil well and conversion of them to a set of indices of 3D cubic structure, mentioned above, might be straightforward as well.

Therefore, for each timestep we have a parallelogram cubic structure, containing petrophysical information about the reservoir, and a set of directives for every oil well in the reservoir. Our goal, given information about the current timestep and all the previous ones, is to predict physical values for all cells for the next timestep. As we have indicated above, straightforward application of conventional machine learning techniques without any preprocessing is infeasible for regular workstations.

To overcome these difficulties, we use a general approach for grid-based Surrogate Reservoir Modeling: to predict physical values for a given cell, we use its state, the state of local neighbor cells, the set of directives for all oil wells for a considered timestep, and geometrical information about the given cell. We will denote examined cell as a target cell. The reasoning behind such approach is that the cell state mostly depends on the state of neighboring cells and the nearest oil wells. The influence of distant cells or oil wells is insignificant if we consider a single timestep. Therefore, such information can be safely disregarded. The cells, which have a common face with considered cell, were chosen as a local set, forming a kind of cross shape. The local set consists of seven cells.

\begin{figure}[t]
\centering
  \includegraphics[scale=0.27]{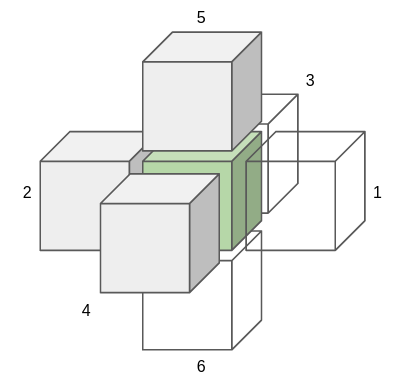}
\caption{Target cell with its local set}
\label{fig:1}
\end{figure}

In Fig. \ref{fig:1} we introduced an enumeration for neighboring cells. Green cell in the center corresponds to the target cell, gray cells are active, and transparent cells are inactive. 

To introduce geometrical information to the algorithm, coordinates of a target cell in parallelogram computational domain and distances to all wells from the target cell were introduced as additional features, as well as running sums of oil extraction, water extraction and injection for all wells. To sum up, a single sample vector contains physical values for a target cell, neighboring cells, oil wells directives, running sums, and geometrical information for the target cell.

As all these values differ in absolute size and have different units of measurement, it is reasonable to normalize them before applying any machine learning algorithm. 

Permeability for all 3 axes and porosity were chosen as static physical characteristics for each cell. Oil saturation and pressure were chosen as dynamic physical characteristics, and will have to be predicted by the resulting algorithm.

\subsection{Surrogate Modeling based on Artificial Neural Networks}

As we are going to predict values for oil saturation and pressure, which, at least explicitly, are governed by different physical laws, we will use a separate ANN for each kind of target variable. Their architecture is almost similar, and the same input values will be used for both networks for each target variable. For initialization and training of ANNs we used algorithms from \cite{Ensembles2013,HDA2013,ANNInit2016}.

As target cells might have different patterns of activity of neighbors ($2^6-1$ neighbors to be exact, excluding the one with all cells in the local set being inactive), one network will be used for each possible pattern. Simply replacing input values corresponding to a given inactive cell with zeros will be not sufficient, as such approximation does not have any implicit physical meaning. This approach might also greatly obscure the calculations. It is convenient to use a dictionary for storing these networks, as each pattern can be represented with 6-digit binary code, where $i$'th position is 0, if cell $i$ is inactive and 1 if it is active. These codes are used as dictionary keys to retrieve neural networks during training.

As we are working with simple numerical data and 3D convolutional neural networks might be impractical for conventional workstations due to the size of the input, standard feedforward architecture will be used for all the networks. In addition, complex high-dimensional structure of the input data, which reflects terrain features of the oil deposit, prevents from using recurrent neural network models.

To sum up, there is a total of $n = 2(2^6-1)$ ANNs, used in learning and prediction processes. Each ANN is designed to process active and neighbouring cells with specific pattern of activity as input and predict one type of dynamic physical characteristic. The number of ANNs is relatively high, but such quantity provides the flexibility, needed for the group of networks to correctly process more peculiarities of the structure of oil deposit for all time steps. For each timestep we iteratively acquire predictions of physical characteristics for each cell using ensemble of ANNs. Prediction for an arbitrary number of timesteps is acquired by sequentially processing individual ones. The outputed dynamic physical characteristics are used as input for the next timestep. Since $n \gg 1$, we used rather simple network architecture to speed up learning process. Each network contains five hidden layers with varying numbers of hidden units. This number of layers was chosen to facilitate fast training and inference even on machines without GPUs without significantly sacrificing representational power. All layers use rectified linear unit as activation function, except for the final layer of saturation prediction model. As saturation values belong to $[0, 1]$ interval, sigmoid activation function allows to effectively restrict output to provide only acceptable saturation values.

\subsection{Model training and prediction routines}

The sheer amount of target cells and neighbors configurations, paired with corresponding geometrical and well information, effectively prevents storing all training set directly. Another procedure needs to be implemented in order to perform training and prediction using described group of ANNs.

	For convenience, each parallelogram computational domain for a given timestep is padded with inactive cells. For each previously active cell on a given timestep we determine the pattern of neighboring active cells, their physical values and geometrical information, then flatten and concatenate them. Well directives for a given timestep are also added. After that, the resulting vector is stored in a dictionary with a key, corresponding to the observed pattern. State of the target cell on the next step is also stored to serve as an output value for training. As the number of vectors in the dictionary for a given pattern reaches predetermined batch size, two corresponding networks are trained on the assembled dataset. After that, values, corresponding to the used batch, are removed. When the final cell for the timestep has been processed, training is performed on all datasets, still remaining in the dictionary before their removal. The algorithm then proceeds to the next timestep.
    
	We have found out, that normalization of input and target cell values, so that they all have zero mean and unit variance among all timesteps, greatly improves prediction quality. Normalizing values only for a given timestep is not enough, as the same normalized values will correspond to different values on the previous timesteps. This preprocessing takes $O(n)$ time, requires constant memory to store means and variances of values, and can easily be reversed. If input or target cell value already belongs to $[0, 1]$ interval, such as saturation for example, no normalization is required.
    
	Both models are trained using the Adam optimizer with default parameters. All the remaining parameters were set to their default values. Pressure prediction model uses $L_2$-loss function, and saturation prediction model is trained with binary cross entropy loss. Using binary cross entropy loss is counterintuitive, but provides better results in practice.
    
	Prediction is done in a similar fashion: for each cell, activity pattern of its neighbors is examined, their values are flattened, concatenated and passed as an input to trained network. By processing all currently active cells we estimate the state of the reservoir on the next step, and by consequently repeating the procedure we recover the state of the reservoir on all examined timesteps.

\section{Results}
\label{sec2}

\begin{figure}[h!]
  \centering
  \centering
\includegraphics[width=1.0\textwidth]{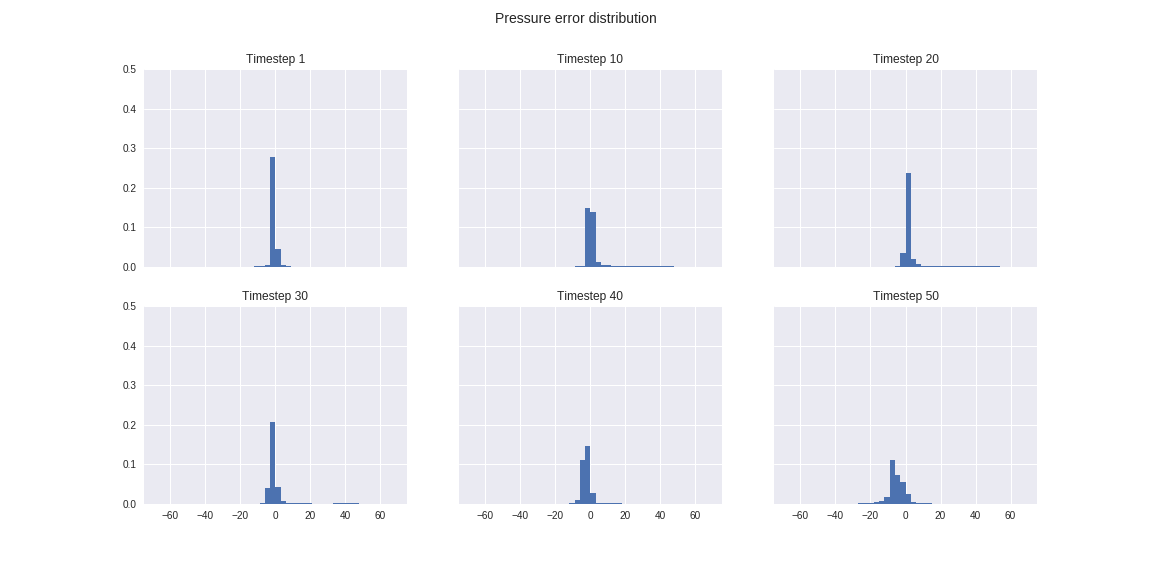}
%
  \caption{Histograms of errors in pressure prediction}
  \label{fig:2}
\end{figure}

\begin{figure}[h!]
\centering
\includegraphics[width=1.0\textwidth]{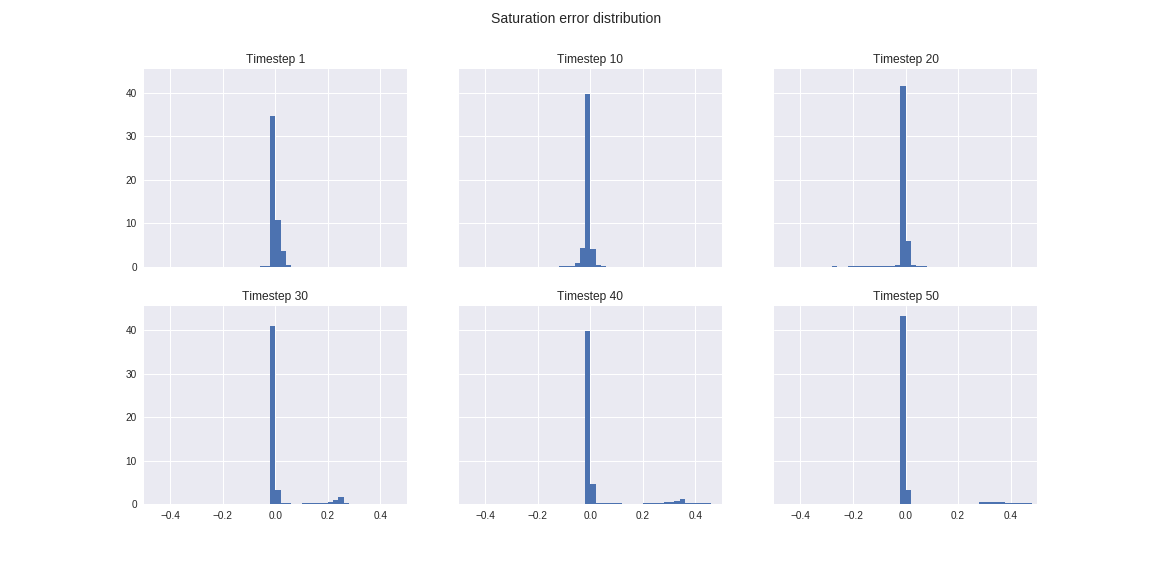}
  \caption{Histograms of errors in saturation prediction}
  \label{fig:3}
\end{figure}

\begin{figure}[h!]
\centering
\includegraphics[width=1.0\textwidth]{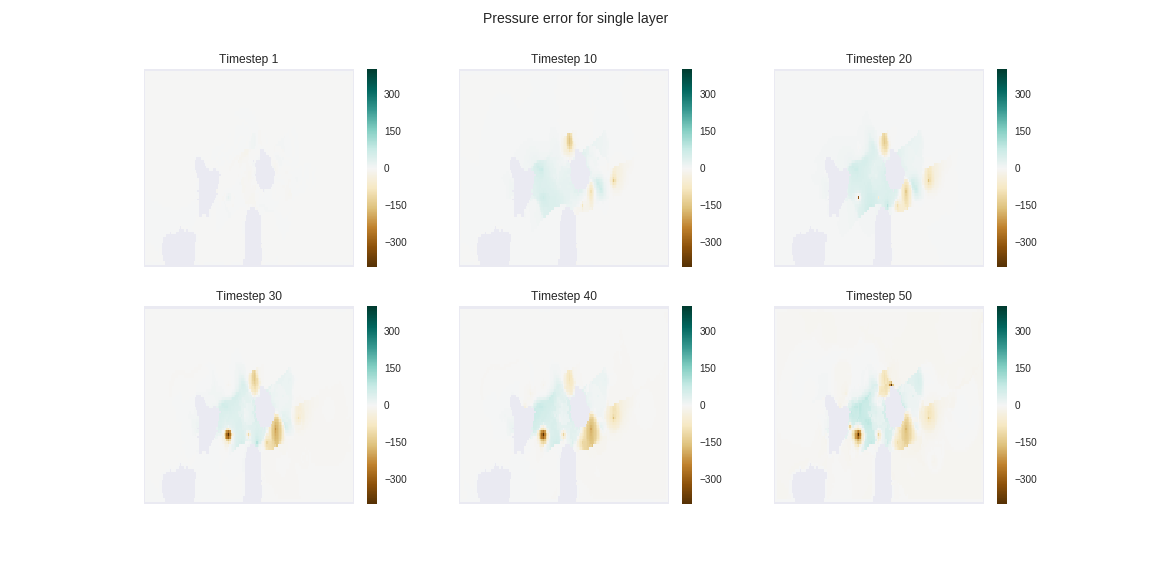}
  \caption{Pressure prediction error for layer 33}
  \label{fig:4}
\end{figure}

\begin{figure}[h!]
\centering
\includegraphics[width=1.0\textwidth]{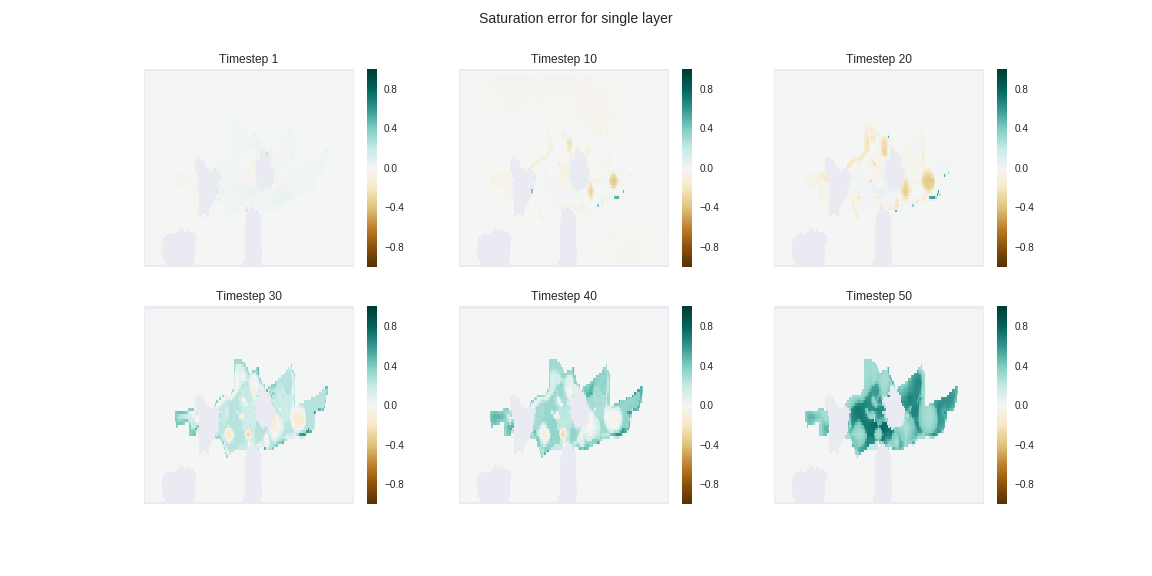}
  \caption{Saturation prediction error for layer 33}
  \label{fig:5}
\end{figure}

We used data from currently developed oil reservoir with size of $70 \times 145 \times 114$ in order to evaluate models’ performance. 20 timesteps, starting from step 0, were used to train the models and $50$ timesteps, starting from the same step, were used to evaluate its performance.

Below are histograms of error distributions for pressure and saturation, see Fig. \ref{fig:2} and Fig. \ref{fig:3}. Predictions for timesteps were obtained sequentially: starting from timestep 0, we use its values as an input to predict values for timestep 1, then we use timestep 1 values to predict timestep 2 values and so on. Note, that the histograms are normalized.

Not all observations are included in the plot. Insignificant number of strong outliers was excluded as abscissa axis was limited to interval $[-75, 75]$, where the majority of observations are located. Error for the most of cells on timesteps from 1 to 20, on which the models were trained, stays close to 0. And, the models were able to provide reasonable estimations for a significant amount of cell configurations on timesteps it had not seen during training. This leads to conclusion that chosen approach  generalizes well and could be used for other applications.

The following set of error heatmaps for the single layer for different timesteps illustrates the same idea: difference between prediction and true value increases in time, see Fig. \ref{fig:4} and Fig. \ref{fig:5}. For all layers, error clusters in a subset of cubes, corresponding to geologically heterogeneous regions, which might have peculiarities not covered by a chosen subset of physical characteristics.

Prediction was done on conventional workstation using NVIDIA GTX 970 GPU. Inference for a single timestep, containing 715105 active cells, requires approximately 104.87344 seconds, or 0.00015 seconds per cell.

\section{Conclusions and Discussion}
\label{sec3}

This paper illustrates that surrogate modeling is a powerful tool for acceleration of routine simulations representing a critical element of field development planning workflow. The area of applicability of the presented approach requires re-training of the data-driven model for each new reservoir which might be considered as a limitation. However, the tool is of practical use for a very fast scenario modeling of reservoirs, for which there exist a history of conventional models usage. The computational performance enables very efficient optimization of field development schemes aimed at minimizing financial risks of cost-intensive decisions on reservoir development and redevelopment planning. 

One potential direction of algorithm improvement includes adding more types of values to the model. Besides simply increasing the amount of value types predicted, one can also improve the quality of inference by employing data augmentation. Data augmentation is a method, mostly used in Computer Vision. It is used to increase the size of the dataset by making use of possible invariances derived from the nature of the dataset samples. For example, if our goal is to recognize, whether the image contains a dog, we can train the model not just on the initial image, but on the rotated, mirrored, and slightly noised images also.

Finally, one could consider to apply adaptive design of experiments, devised for industrial engineering problems, to simultaneously increase efficiency of sensitivity analysis and to improve utilization of computational budget of generating a training sample \citep{burnaev2017efficient,burnaev2015adaptive,ADoESobol2015,ADoESobol2016}.

As both training and prediction are done for all cubes independently, one may argue that using a distributed programming model, such as MapReduce can be a viable choice. Implementation of both map and reduce phases both for training and prediction is straightforward, but its description goes outside of the scope of this article.

\section*{Acknowledgements}

The work was supported by the Russian Science Foundation under Grant 19-41-04109.

\bibliographystyle{splncs04}
\bibliography{mybibfile2}

\end{document}